\documentclass[Times,4pt,aps,pra,onecolumn,showpacs,amsmath,amssymb,floatfix,footinbib,superscriptaddress]{revtex4-1}
\usepackage{amsmath,natbib,graphicx,subfigure,amssymb,graphics,amsmath,mathrsfs,CJK,color}
\usepackage{multirow,fancyhdr,color,bm,tabularx,psfrag,dcolumn}
\usepackage[colorlinks=true,linkcolor=blue,urlcolor=blue,citecolor=blue]{hyperref}
\usepackage[left]{lineno}
\usepackage{tikz,xcolor,hyperref}
\definecolor{lime}{HTML}{A6CE39}
\DeclareRobustCommand{\orcidicon}{%
    \begin{tikzpicture}
    \draw[lime, fill=lime] (0,0)
    circle [radius=0.16]
    node[white] {{\fontfamily{qag}\selectfont \tiny ID}};\draw[white, fill=white] (-0.0625,0.095)
    circle [radius=0.007];
    \end{tikzpicture}
    \hspace{-2mm}}
\foreach \x in {A, ..., Z}
{\expandafter\xdef\csname orcid\x\endcsname{\noexpand\href{https://orcid.org/\csname orcidauthor\x\endcsname}{\noexpand\orcidicon}}}

\begin{document}

\title{ Photovoltaic performances in a cavity-coupled double quantum dots photocell }
\author{Sheng-Qiang Zhong}
\affiliation{Department of Physics, Faculty of Science, Kunming University of Science and Technology, Kunming, 650500, PR China}

\author{Shun-Cai Zhao \orcidA{}}
\email[Corresponding author: ]{zhaosc@kmust.edu.cn}
\affiliation{Department of Physics, Faculty of Science, Kunming University of Science and Technology, Kunming, 650500, PR China}

\author{Sheng-Nan Zhu }
\affiliation{Department of Physics, Faculty of Science, Kunming University of Science and Technology, Kunming, 650500, PR China}

\begin{abstract}
Revealing the quantum regime of photovoltaics is crucial to enhancing the internal quantum efficiency of a double quantum dots (DQDs) photocell housed in a cavity. In this study, the performance of a quantum photovoltaic is evaluated based on the current-voltage and power-voltage characteristics in a cavity-coupled DQDs photocell. The results show that the cavity-DQDs coupling coefficient plays a dissipative role in the photovoltaic performance, and the cavity has a limited size for the photovoltaic performance. Additionally, more low-energy photons are easily absorbed by this cavity-coupled DQDs photocell compared with the case without cavity. These results may provide some strategies for improving the photoelectric conversion efficiency and internal quantum efficiency of cavity-coupled DQDs photocells.
\begin{description}
\item[PACS numbers]42.50.Gy
\item[Keywords]{Cavity-coupled double quantum dots photocell, cavity-DQDs coupling coefficient, low-energy photons}
\end{description}
\end{abstract}
\maketitle
\section{Introduction}

Double quantum dots (DQDs) have been used for many different purposes in recent decades due to their small size and adjustable band-gap energy \cite{1998PhyB..249..206A,Ding2002ElectrochemistryAE,doi:10.1063/1.3562192,2003Natur.423..422L,AMAHA2001183}.
In some cases, DQDs are regarded as artificial molecules \cite{1998PhyB..249..206A} owing to their counterintuitive physical properties, such as broad absorption with narrow photoluminescence spectra \cite{doi:10.1063/1.3562192,Ding2002ElectrochemistryAE}, quantum-tunneling effect \cite{2003Natur.423..422L,AMAHA2001183}, and low photobleaching and resistance to chemical degradation \cite{1999PhRvB..59.5688S}. Compared with the single QD, DQDs are more suitable to process various quantum effects in complex environments \cite{Fujisawa2019Spontaneous,AMAHA2001183,2000PhRvL..85.1946A,2001Sci...291..451B,2002Sci...297.1313O,PhysRevLett.90.026602} due to their larger bandwidth and lower noise detection \cite{PhysRevLett.108.046807}.

Recent studies have illustrated that the photoelectric conversion efficiency can be greatly enhanced by the QD photocell \cite{2010Quantum,Dorfman2746,2011Peak,2013Optimal,Zhao2019,2020High}. In the multi-band QD photocell \cite{Zhao2019} scheme, the high current density keeps the output voltage unaffected while more low-energy photons are absorbed, leading to a greater output efficiency in the multi-band QD photocell.
Quantum coherence has also been proven to play a role in semiconductor QDs \cite{2007Analytical} and heterostructures \cite{PhysRevA.63.053803}. Recently, it was demonstrated that quantum coherence induced by an external source can increase the photocell output power \cite{2010Quantum}. Our previous work \cite{ZHONG2021104094} demonstrated that the electron tunneling effect between two QDs in the DQDs photocell leads to the redistribution of populations on two QDs, which ultimately leads to the improvement of the photovoltaic properties of the system. In addition, quantum coherence has been demonstrated to modify photon absorption and emission \cite{2010Quantum,Dorfman2746}, such as lasing without inversion \cite{1996Electromagnetically}, electromagnetically induced transparency \cite{2009Plasmonic}, and slow light \cite{0Light} in atomic systems.

Considering the internal quantum efficiency caused by the packaging technique in actual photocells, we established a cavity-coupled DQDs photocell model and evaluated its photovoltaic performance without considering the Coulomb-interaction between different electrons\cite{PhysRevB.85.075306,2013Stepwise}. The photovoltaic performance was measured by some parameters, such as the coupling-coefficient between the cavity and DQDs photocell system, the size of the cavity, and the absorbed-photon wavelength.

\section{ Model and equations}

\subsection{ Hamiltonian of the cavity-coupled DQDs photocell }

\begin{figure}[htp]
\center
\subfigure[]{\includegraphics[width=1.8in]{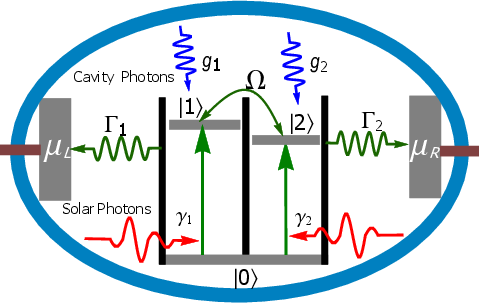}}
\hspace{0.20in}
\subfigure[]{\includegraphics[width=1.6in]{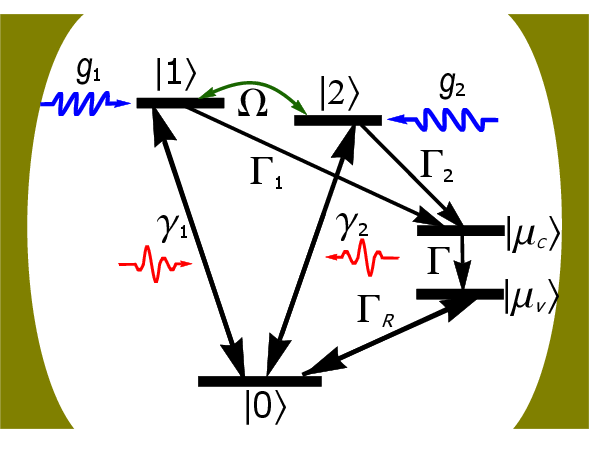}}
\caption{(Color online) (a) Sketch of the housed DQDs photovoltaic cell. The horizontal gray solid lines represent the electronic states of the DQDs, and electrons are confined to the L (\emph{L}=1) and R (\emph{R}=2) dots by barrier gates (black vertical solid lines). When the sun sheds light on the surface of the microcavity, electronic transport is driven in the L and R dots by the absorbed-photons (Red wavy lines) at the rates $\gamma_{i} (i=1,2)$, and the electrons are released at the rates \(\Gamma_{1}\), \(\Gamma_{2}\) from the left/right electron leads (with their chemical potentials $\mu_{L}$, $\mu_{R}$). Tunneling rates \(\Omega\) between the dots can be tuned via gate-controlled tunnel barriers. The coupling between the microcavity photons (wavy blue lines) and each QD is described by a Holstein interaction of strength \(g_{1}\), \(g_{2}\). (b) Energy-level schematic of the microcavity-coupled DQDs photocell corresponding to (a): the solar photon absorption assists the excitation (relaxation) between the eigen-state $|i\rangle_{(i=1,2)}$ $\leftrightarrow$ $|0\rangle$ by $\gamma_{i(i=1,2)}$. The relaxations $\Gamma_{i(i=1,2)}$ between $|i\rangle_{(i=1,2)}$ and $|\mu_{c}\rangle$ are accompanied by electron transport from the DQDs to the two electronic leads. Levels $|\mu_{c}\rangle$ and $|\mu_{v}\rangle$ are connected to an output at the rate $\Gamma$. Transitions $|i\rangle_{(i=1,2)}$ $\leftrightarrow$ $|\mu_{c}\rangle$, $|\mu_{v}\rangle$ $\leftrightarrow$ $|0\rangle$ are driven by ambient thermal phonons.}
\label{f1}
\end{figure}

The proposed microcavity coupled DQDs photocell sketch is shown in Fig. \ref{f1}(a), which is housed in a microcavity by the Coulomb blockade regime. This indicates that the DQDs are restricted to three possible configurations, namely, the null-electron subspace, denoted by $|0\rangle$, and the single-electron subspace with an electron localized either on the left or right dot, denoted by $|1\rangle$ and $|2\rangle$, respectively. We set the energy of the unoccupied electronic state at zero. The DQDs photocell system is initiated by the absorbed photons at the rates $\gamma_{1}$ and $\gamma_{2}$, respectively.

After the absorption of cavity photons, electrons are transported via the tunneling coefficient $\Omega$ between two states $|1\rangle$ and $|2\rangle$, which can be flexibly tuned via gate voltages applied on the dots \cite{2002Electron,Zhao2009}. Each dot is further coupled to the cavity mode with their coupling-strength denoted by \(g_{1}\) or \(g_{2}\). The two QDs are placed between two metal leads composed of noninteracting electrons, which act as the fermionic reservoirs with the chemical potentials $\mu_{L}$ and $\mu_{R}$. Therefore, electron transfer between two dots and two electric leads occurs via direct tunneling described by tunneling rate \(\Gamma_{i (i=1,2)}\), which can also be tuned via the external gate voltages applied on the dots \cite{2002Electron,Zhao2009}. Ultimately, the electric energy is collected by the connected external output terminal.

In the process of the sun shedding light on the cavity surface, cavity gain occurs, and the cavity photons couple with the double quantum dots (see Fig.\ref{f1}). The loss of cavity-coupled DQDs is taken into account via the cavity-DQDs coupling coefficient $\kappa$ in this proposed photocell model. Therefore, the total DQDs photocell system can be modeled by the Hamiltonian \cite{Fujisawa2019Spontaneous} as follows,

\vspace{0.25pt}
\begin{equation}
\hat{H}_{T}=\hat{H}_{D}+\hat{H}_{ph}+\hat{H}_{D-ph},
\end{equation}

\noindent where \(\hat{H}_{D}\) describes the electronic properties in the DQDs photocell system with

\begin{equation}
\hat{H}_{D}=\frac{\varepsilon}{2}\hat{\sigma}_{z}+\Omega \hat{\sigma}_{x}+\sum_{k}\omega_{k}\hat{C}_{k}^ {\dag}\hat{C}_{k}+(\sum_{k}\sum_{i=1,2}g_{ki}\hat{d}_{i}^ {\dag}\hat{C}_{k}+H.c.),
\end{equation}

\noindent where $\varepsilon$ is the electrostatic energy mismatch between the DQDs, and $\Omega$ is the inter-dot tunnel coupling coefficient \cite{2013Quantum}. The Pauli operators are defined as $\hat{\sigma}_{z}$= $\hat{d}_{L}^{\dag}$$\hat{d}_{L}$ $-$ $\hat{d}_{R}^ {\dag}$$\hat{d}_{R}$, $\hat{\sigma}_{x}$= $\hat{d}_{L}^ {\dag}$$\hat{d}_{R}$ + $\hat{d}_{R}^ {\dag}$$\hat{d}_{L}$, with $\hat{d}_{i}^ {\dag}$($\hat{d}_{i}$)($i$=1,2) respectively representing the creation (annihilation) operators of an electron in the left or right QD. The electrodes are modeled as collections of noninteracting electrons via the creation (annihilation) operator $\hat{C}_{k}^ {\dag}$($\hat{C}_{k}$) for an electron with momentum \(k\). The tunnel-coupling term between the $i$th ($i$=1,2) dot and its corresponding lead in Eq. (2) is described by $\sum\limits_{k}\sum\limits_{i=1,2}g_{k i}\hat{d}_{i}^ {\dag}\hat{C}_{k}+H.c.$ with the coupling constant $g_{k i}$, and \(H.c.\) denotes the Hermitian conjugate terms of the previous parts.

In Eq. (2), the term \(\hat{H}_{ph}\) denotes the cavity photons interacting with the cavity in the form of standing-waves with the following Hamiltonian,

\begin{equation}
\hat{H}_{ph}=\hbar\omega_{0}\hat{a}^{\dag}\hat{a}+2\hbar A (\hat{a}^{\dag}+\hat{a})cos\omega t,
\end{equation}

\noindent where \(\hat{a}\)(\(\hat{a}^{\dag}\)) denotes the annihilation (creation) operators for photons in the cavity, \(A\) is the amplitude of the external drive of the cavity, and \(\omega\) is the
frequency of the source. The interaction between the microwave field and the DQDs system is represented by

\begin{equation}
\hat{H}_{D-ph}=\hbar\sum_{i=1,2} g_i(\hat{a}^{\dag}+\hat{a})\hat{\sigma}_{z},
\end{equation}

\noindent where $g_{i(i=1,2)}$ denotes the corresponding coupling-constant to the corresponding QD.

\subsection{ Master equations for the cavity-coupled DQDs photocell}

After completing the above deduction of the Hamiltonian for this DQDs photocell system, the Born-Markov second order master equation takes the following form in the Schr$\ddot{o}$dinger picture,

\begin{equation}
\frac{\partial}{\partial t}\hat{\rho}(t)=-i[\hat{H}_{D},\hat{\rho}]+\hat{{\cal L}}_{i} \hat{\rho}+\hat{{\mathcal{L}}}_{\Gamma_i}\hat{\rho} +\hat{{\mathcal{L}}}_{\Gamma_R}\hat{\rho}+\hat{{\mathcal{L}}}_{\Gamma}\hat{\rho}+\hat{{\mathcal{L}}}_{\kappa}\hat{\rho},
\end{equation}

The superoperator $\hat{\mathcal{L}}$ is decomposed into parts for describing the dissipative behaviors between the photovoltaic system and the cavity with $\hat{\cal L}_{i}\hat{\rho}, \hat{{\mathcal{L}}}_{\Gamma_i}\hat{\rho}, \hat{{\mathcal{L}}}_{\Gamma_R}\hat{\rho}, \hat{{\mathcal{L}}}_{\Gamma}\hat{\rho}, \hat{{\mathcal{L}}}_{\kappa}\hat{\rho}$. The dissipative behaviors between the DQDs photovoltaic system and the ambient environment in the cavity can be read as,

\begin{equation}
\hat{\mathcal{L}}_{i}\hat{\rho} = \sum_{i=1,2}\frac{\gamma_{i}}{2}[ (n_{i}+1)\mathcal{D}[\hat{\sigma}_{0i}]\hat{\rho}+n_{i}\mathcal{D}[\hat{\sigma}_{0i}^{\dag}]\hat{\rho}],
\end{equation}

\noindent where $n_{i}$  is the average number of photons and is expressed as $n_{i} = \frac{1}{exp[\frac{E_{i0}}{k_B T_s}]-1}$, which describes the number of solar photons absorbed by the DQDs system according to the sun temperature $T_s$. Moreover, the specific expression of the mark $\mathcal{D}$ acting on any operator \(\hat F\) is defined as $\mathcal{D}[ \hat F ]\hat\rho=2\hat F\hat\rho\hat F^{\dag}-\hat\rho\hat F^{\dag}\hat F-\hat F^{\dag}\hat F\hat\rho$. Superoperator $\hat{\mathcal{L}}_{\Gamma_i}\hat{\rho}$ corresponding to the quantum coherence from the dot-lead coupling can be described by

\begin{equation}
\hat{\mathcal{L}}_{\Gamma_i}\hat{\rho} = \sum_{i=1,2}\frac{\Gamma_i}{2}[ (n_{ic}+1)\mathcal{D}[\hat{\sigma}_{\mu_c i}]\hat{\rho}+n_{ic}\mathcal{D}[\hat{\sigma}_{\mu_c i}^{\dag}]\hat{\rho}],
\end{equation}

\noindent where $\hat{\sigma}_{\mu_c i} = |\mu_c\rangle\langle i|$ (i=1,2), and the corresponding average phonon number is $n_{ic} = \frac{1}{exp[\frac{E_{i\alpha}}{k_B T_a}]-1}$ with the ambient temperature $T_a$. $\Gamma_i$ denotes the electronic transport from (to) the level $|i\rangle$ to (from) the level $|\mu_c\rangle$. Similarly, the dissipative process between the drain ($|\mu_v\rangle$) and the ground state ($|0\rangle$) can be written as,

\begin{equation}
\hat{\mathcal{L}}_{\Gamma_R}\hat{\rho} = \frac{\Gamma_R}{2}[ (N_{c}+1)\mathcal{D}[\hat{\sigma}_{0\mu_v}]\hat{\rho}+N_{c}\mathcal{D}[\hat{\sigma}_{0\mu_v}^{\dag}]\hat{\rho}],
\end{equation}

\noindent Here, we use states $|\mu_c\rangle$ and $|\mu_v\rangle$ to represent the source and drain ($\mu_R$, $\mu_L$), respectively [as shown in Fig.\ref{f1}(b)]. Additionally, $\hat{\sigma}_{0\mu_v} = |0\rangle\langle \mu_v|$, the average number of phonons  is $N_{c} = \frac{1}{exp[\frac{E_{0\beta}}{k_B T_a}]-1}$, and $\Gamma_R$ is the spontaneous decay rate from the level $|\mu_v\rangle$ to the ground state $|0\rangle$.

The dissipative process between the cavity and the DQDs is denoted as,

 \begin{equation}
 \hat{{\mathcal{L}}}_{\kappa}\hat{\rho} = \frac{\kappa}{2}(2\hat{a}\hat{\rho}\hat{a}^{\dag}-\hat{a}^{\dag}\hat{a}\hat{\rho}-\hat{\rho}\hat{a}^{\dag}\hat{a})
 \end{equation}

\noindent where $\hat{a}^ {\dag}$($\hat{a}$) depicts the creation (annihilation) operator for an electron in the cavity-coupled DQDs photocell system via the cavity-DQDs coupling coefficient $\kappa$, which is a function of the coupling-constant $g_{i(i=1,2)}$ \cite{2017Quantum}. For simplicity, we treat $g_{1}$ and $g_{2}$ as equal quantities in the following calculation process. Finally, a process with relaxation rates $\Gamma$ proportional to the output electronic current is defined for the system decaying from state $|\mu_c\rangle$ to state $|\mu_v\rangle$ as follows,

\begin{equation}
\hat{{\mathcal{L}}}_{\Gamma}\hat{\rho} = \frac{\Gamma}{2}(2|\mu_v\rangle\langle \mu_c|\hat\rho|\mu_c\rangle\langle \mu_v|-|\mu_c\rangle\langle \mu_c|\hat\rho-\hat\rho|\mu_c\rangle\langle \mu_c|).
\end{equation}

\begin{table}[htp]
\begin{center}
\caption{Parameters utilized for this cavity-coupled DQDs photocell system.}
\label{T1}
\setlength{\tabcolsep}{5mm}{
\begin{tabular}{cccc}
\hline
\hline
         & \(parameters\)        & \( Units \)   & \(Values\)          \\
\hline
         & $\Gamma$              & eV            &0.12                 \\
         & $\Gamma_{R}$          & eV            & 0.024               \\
         & $\Gamma_{1}$          & eV            & 0.14                \\
         & $\Gamma_{2}$          & eV            & 0.02                \\
         & $\gamma_{1}$          & eV            & $6.20*10^{-7}$      \\
         & $\gamma_{2}$          & eV            & $1.98*10^{-7}$      \\
         & $T_a$                 & K             & 300                 \\
         & $T_s$                 & K             & 5800                \\
         & $n_{ic}$              &               & $5.98*10^{-4}$      \\
         & $N_{c}$               &               & $4.57*10^{-4}$      \\
         & $k_B$                 &               & 1                   \\
\hline
\hline
\end{tabular}}
\end{center}
\end{table}

\subsection{Steady-state photovoltaic performance}

\par Under the Weisskopf-Wigner approximation \cite{1974On}, we can obtain the dynamic equations of the corresponding matrix elements for this proposed photocell system as follows,

\begin{widetext}
\begin{eqnarray}
&\dot{\rho}_{11}=&-i\Omega(\rho_{21}-\rho_{12})-\kappa\rho_{11}-\Gamma_{1}[(n_{1c}+1)\rho_{11}-n_{1c}\rho_{\mu_c\mu_c}]-\gamma_{1}[(n_{1}+1)\rho_{11}-n_{1}\rho_{00}],\nonumber\\
&\dot{\rho}_{22}=&i\Omega(\rho_{21}-\rho_{12})-\kappa\rho_{22}-\Gamma_{2}[(n_{2c}+1)\rho_{22}-n_{2c}\rho_{\mu_c\mu_c}]-\gamma_{2}[(n_{2}+1)\rho_{22}-n_{2}\rho_{00}],\nonumber\\
&\dot{\rho}_{\mu_c\mu_c}=&\Gamma_{1}[(n_{1c}+1)\rho_{11}-n_{1c}\rho_{\mu_c\mu_c}]+\Gamma_{2}[(n_{2c}+1)\rho_{22}-n_{2c}\rho_{\mu_c\mu_c}]-\Gamma\rho_{\mu_c\mu_c},\nonumber\\
&\dot{\rho}_{\mu_v\mu_v}=&\Gamma\rho_{\mu_c\mu_c}-\Gamma_{c}[(N_{c}+1)\rho_{\mu_v\mu_v}-N_{c}\rho_{00}],\\
&\dot{\rho}_{12}=&-i(\varepsilon-\omega)\rho_{12}-i\Omega(\rho_{22}-\rho_{11})-\kappa\rho_{12}-\frac{\rho_{12}}{2}[\gamma_{1}(n_{1}+1)+\gamma_{2}(n_{2}+1)+\Gamma_{1}(n_{1c}+1)\nonumber\\&&+\Gamma_{2}(n_{2c}+1)],\nonumber\\
&\dot{\rho}_{21}=&i(\varepsilon-\omega)\rho_{21}+i\Omega(\rho_{22}-\rho_{11})-\kappa\rho_{21}-\frac{\rho_{21}}{2}[\gamma_{1}(n_{1}+1)+\gamma_{2}(n_{2}+1)+\Gamma_{1}(n_{1c}+1)\nonumber\\&&+\Gamma_{2}(n_{2c}+1)].\nonumber
\end{eqnarray}
\end{widetext}

Thus, we focus on the steady-state photovoltaic characteristics of the cavity-coupled DQDs photocell. As mentioned in the DQDs photocell model \cite{1965CONTRIBUTIONS}, the generated current is considered to flow through the load connected to the acceptor $|\mu_c\rangle$ and $|\mu_v\rangle$, and the effective voltage $V$ is defined as $eV=\mu_{c}-\mu_{v}+k_BT_a\ln(\frac{\rho_{\mu_c\mu_c}}{\rho_{\mu_v\mu_v}})$ \cite{2010Quantum}, a drop of the electrostatic potential across the external load. The current formed in the electron transport process is the variation of particle number with time, so the current tunneling through the source-drain bias ($\mu_{c},\mu_{v}$) can be defined as $j=e\frac{dN(t)}{dt} $ \cite{1999Shot}, where N(t) is the number of electrons arriving at the electron reservoir at time t. Therefore,
the effective electric current formed in this transport process is equivalently written as $j= e \Gamma{\rho}_{\mu_c\mu_c}$. With the steady solution to Eq. (11), the photovoltaic performance of this cavity-coupled DQDs photocell can be evaluated.

\section{Results and discussions}

\begin{figure}[htp]
\center
\includegraphics[totalheight=2 in]{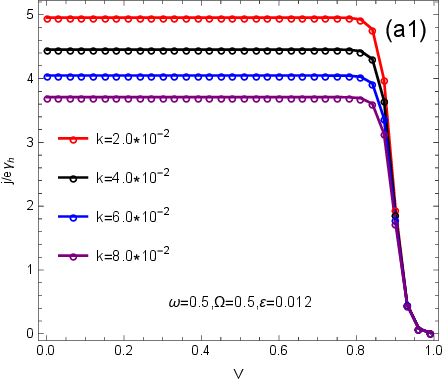 }\includegraphics[totalheight=2 in]{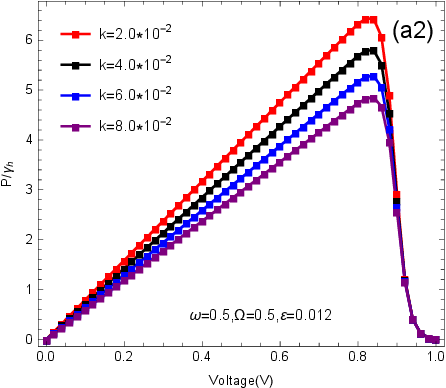 }
\caption{$\mathbf{}$ (Color online) (a1) Current-voltage and (a2) power-voltage characteristics with different ccavity-DQDs coupling coefficients $\kappa$ at room temperature $T_a$ = 300 K. Other parameters are taken from Table \ref{T1}.}\label{p2}
\end{figure}

In the proposed cavity-coupled DQDs photocell system, the parameter $\omega$ describing the features of the cavity and the cavity-DQDs coupling coefficient $\kappa$ between the cavity and DQDs system are the most attractive parameters because they indicate the difference between the DQDs photocell \cite{ZHONG2021104094} and the cavity-coupled DQDs photocell. Other selected parameters \cite{2016A,2016Vibration,Wertnik2018Optimizing} for the cavity-coupled DQDs photocell model are listed in Table \ref{T1}. Therefore, the photovoltaic properties dependent on the cavity-DQDs coupling coefficient $\kappa$ are shown in Fig. \ref{p2} by the current-voltage and power-voltage.

The current-voltage characteristics in Fig.\ref{p2}(a1) indicate the negative role of $\kappa$ in the output short-circuit current, which can be discovered by the curves in Fig. \ref{p2}(a1). The shift from the red curve to the purple curve illustrates that the short-circuit currents decrease with the increase of the cavity-DQDs coupling coefficient $\kappa$ from $2\times10^{-2}$ to $8\times10^{-2}$. Similarly, the cavity-DQDs coupling coefficient $\kappa$ has a passive effect on the power-voltage characteristic, which can be concluded from the curves in Fig. \ref{p2}(a2) showing the reduced peak powers with the increase of $\kappa$. Comparing the photovoltaic properties in this photocell model with the laser behavior in a cavity, it may be concluded that the dissipation caused by the cavity has a similar physical regime.

\begin{figure}[htp]
\center
\includegraphics[totalheight=2 in]{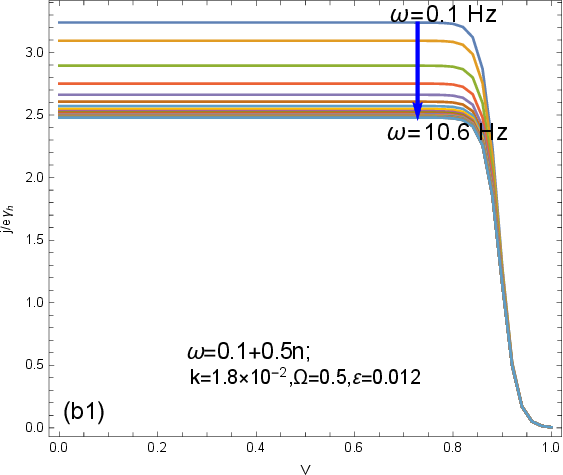 }\includegraphics[totalheight=2 in]{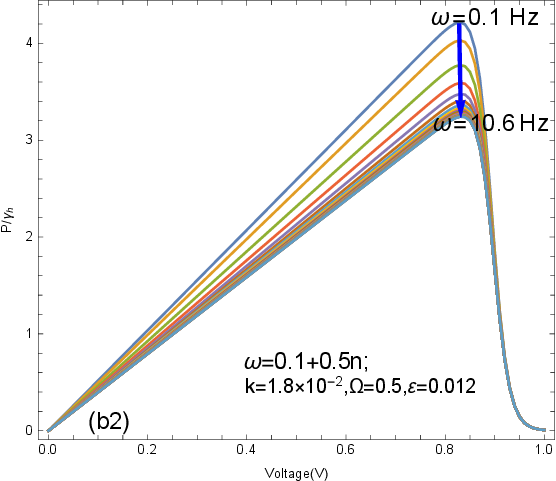 }
\caption{(Color online) (b1) Current-voltage and (b2) power-voltage characteristics with different $\omega$. Other parameters are the same as in Fig. \ref{p2}.}
\label{p3}
\end{figure}

Owing to the general physical formula $\omega$=$\frac{2\pi c}{L}$, where \textit{c} is the velocity and \textit{L} is the length of traveling wave, the size can be denoted by the length of traveling wave in the cavity. Therefore, the parameter $\omega$ from the external microwave source can be indirectly used to describe the features of the cavity. Adjusting the value of $\omega$ can manipulate the size of the cavity in this DQDs photocell system. Next, we change the size of the cavity according to the formula $\omega=0.1+0.5n$. The directions of the two blue arrows in Fig. \ref{p3} (b1) and (b2) both intuitively indicate the influence of $\omega$ on the photovoltaic performance. As shown in Fig. \ref{p3}, the current-voltage curves and peak power curves keep decreasing with the increment of $\omega$ from 0.1 Hz to 10.6 Hz. Moreover, we notice that in the subsequent increment of $\omega$, the decrements in the short-circuit current and peak output power are getting smaller and smaller. Finally, both of them almost reach a stable minimum. Due to the inverse proportional relationship between $\omega$ and $L$, the above results imply that poorer photovoltaic performance is caused by a smaller cavity. That is to say, there is a limit to the size of the cavity for the photovoltaic performance.

\begin{figure}
\center
\includegraphics[totalheight=2 in]{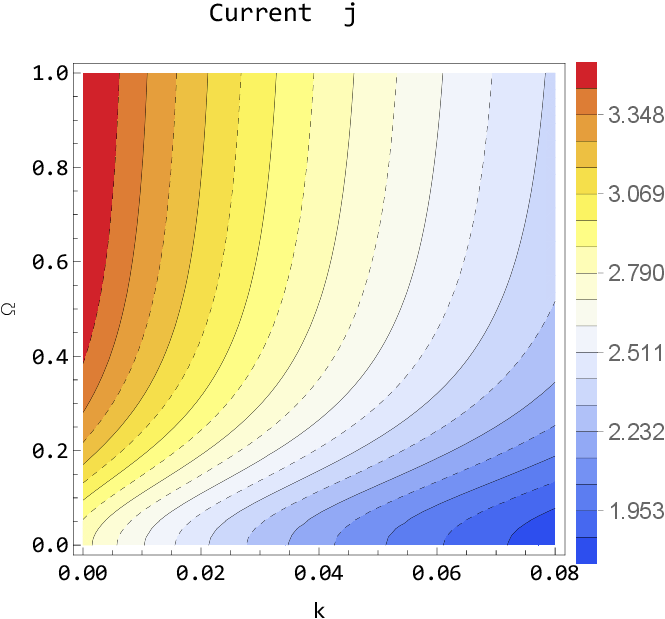 }\includegraphics[totalheight=2 in]{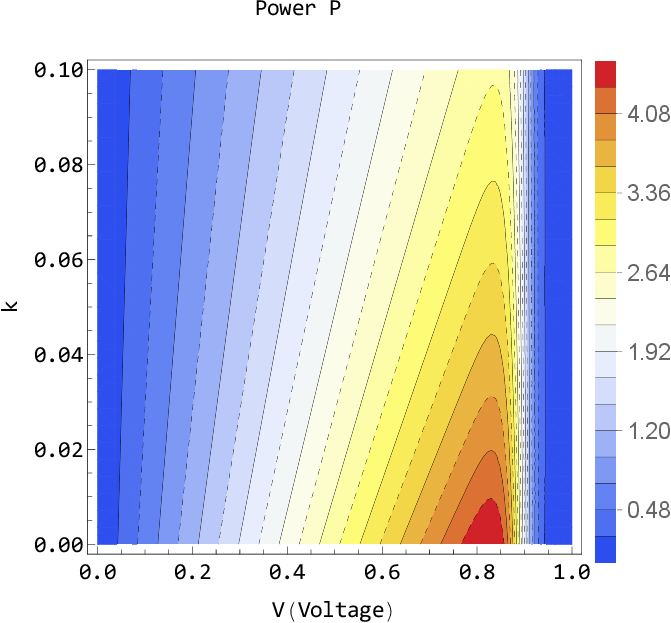 }
\caption{$\mathbf{}$  (Color online) Contour of the photovoltaic current as a function of the cavity-DQDs coupling coefficients $\kappa$ and tunneling coefficients $\Omega$ (top figure). Contour of the output power as a function of $\kappa$ and output voltage (bottom figure). Other parameters are the same as in Fig. \ref{p2}.}
\label{p4}
\end{figure}

Meanwhile, the influence of the inter-dot tunnel coefficient $\Omega$ and voltage on the photovoltaic performance should be revisited due to the cavity in the DQDs photocell system. As shown by the contour plots in Fig. \ref{p4}, the current $j$ increases with the inter-dot tunnel $\Omega$ but decreases with $\kappa$, which can be illustrated by the output $j$ in the red district, with $\Omega$ being approximately in the range [0.4, 1.0]. This is obviously different from the case of the DQDs photocell without the cavity \cite{ZHONG2021104094}. The output power intuitively shows its maximum reduces with the increment of $\kappa$ around 0.82 $V$, which is in full agreement with the results shown in Fig.\ref{p3}(b2). These results demonstrate that other parameters are greatly influenced by $\kappa$ in the cavity-coupled DQDs photocell system.

\begin{figure}[htp]
\center
\includegraphics[totalheight=2 in]{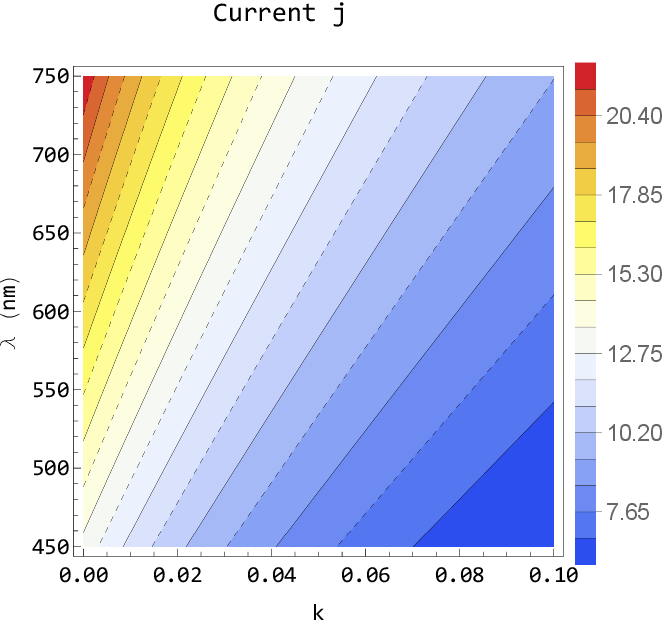 }\includegraphics[totalheight=2 in]{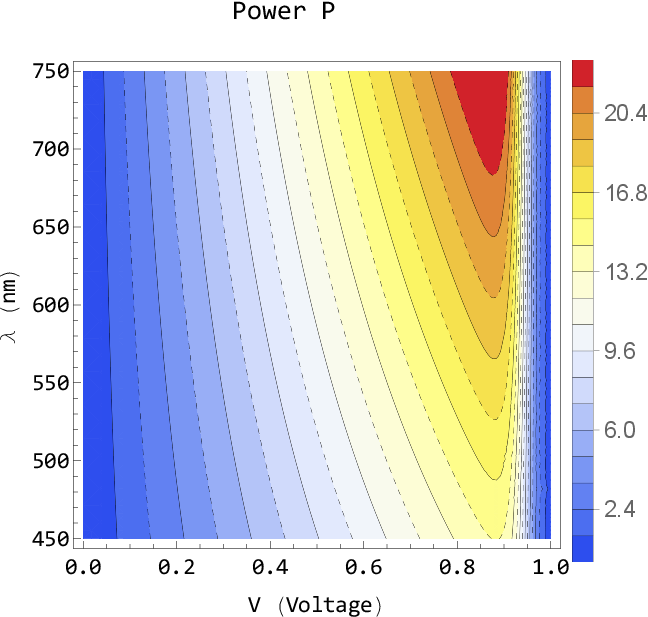 }
\caption{ (Color online) Contour of the photovoltaic current as a function of the absorbed solar photon wavelength $\lambda$ and cavity-DQDs coupling coefficients $\kappa$ (top figure). Contour of the output power (bottom figure) as a function of the absorbed solar photon wavelength $\lambda$ and output voltage with $\kappa$=$2\times10^{-2}$. Other parameters are the same as in Fig. \ref{p2}.}
\label{p5}
\end{figure}

In this cavity-coupled DQDs photocell model, the absorbed photons' wavelength may be changed by the environment in the cavity. Fig. \ref{p5} shows the photovoltaic properties dependent on the the absorbed photons' wavelength $\lambda$ via the photovoltaic current (top figure) and output power (bottom figure). The color changes in different districts in Fig. \ref{p5} similarly reflect the negative effect of $\kappa$, and we also notice that the output photovoltaic current reaches its maximum in the long-wavelength interval, roughly in the range of [700 nm, 750 nm] (see the top figure). Furthermore, the output current in the long-wavelength interval is rapidly weakened by the increase of $\kappa$. Under the condition of $\kappa$=$2\times10^{-2}$, the output power reaches the maximum value around 0.82 V, and the maximum value gradually increases with the increment of the absorbed-photon wavelength. The output power reaches its peak in the wavelength range of 700--750 nm (see the bottom figure). As we all know, the low-energy photons are not absorbed owing to the energy below the corresponding band-gap energy. The above results indicate that low-energy photons are more likely to be absorbed by the cavity-coupled DQDs photocell system. Furthermore, compared with the multi-level QD photocell scheme \cite{Zhao2019}, it is found
that the cavity-coupled DQDs photocell system achieves greater absorption of low-energy photons. Undoubtedly, the above results prove that a greater photoelectric conversion efficiency can be reached by the absorbed lower-energy photons.

Before concluding this section, we need to mention that the photovoltaic performance in the cavity-coupled DQDs photocell is affected by the internal and external quantum efficiencies, and there are many factors that influence the internal quantum efficiency, such as quantum fluctuations, band-gap energies in different materials, and ambient temperature. However, here, we only discuss the features of the cavity and the coupling-parameter between the cavity and DQDs photocell system. Considering the quantum fluctuations and ambient temperature in the DQDs photocell housed in a cavity, it is necessary to establish another DQDs photocell theoretical model. In our follow-up work, we will carry out this task.

\section{Conclusions}

In conclusion, the photovoltaic properties of a cavity-coupled DQDs photocell are explored via the current-voltage and power-voltage characteristics, and the roles of the cavity-DQDs coupling coefficient, the size of the cavity, and the wavelength of absorbed photons are discussed in detail. The results reveal the dissipative influence of the cavity-DQDs coupling coefficient and prove that there is a limit to the size of the cavity for the photovoltaic performance. Furthermore, low-energy photons can be easily absorbed by the cavity-coupled DQDs photocell system, which broadens the absorption photon spectrum of the DQDs photocell system.
The above results offer some strategies to enhance the internal quantum efficiency and photoelectric conversion efficiency of the cavity-coupled DQDs photocell, which may promote the development of photovoltaic devices in the future.

\begin{acknowledgments}
We offer our thanks for the financial support from the National Natural Science Foundation of China (Grant Nos. 62065009 and 61565008 ).
\end{acknowledgments}

\section*{Conflict of Interest}
The authors declare that they have no conflict of interest or personal relationships. This article does not contain any studies with human participants or animals performed by any of the authors. Informed consent was obtained from all individual participants included in the study.





\bibliography{reference}
\bibliographystyle{unsrt}
\end{document}